\documentclass[conference]{IEEEtran}
\IEEEoverridecommandlockouts

\usepackage{cite}
\usepackage{amsmath,amssymb,amsfonts}
\usepackage{graphicx}
\usepackage{textcomp}
\usepackage{xcolor}
\usepackage{booktabs}
\usepackage{algorithm}
\usepackage{amsmath}
\usepackage{algpseudocode}
\usepackage{float}
\usepackage{newfloat}
\usepackage{subcaption}
\usepackage[margin=1in]{geometry}
\usepackage{url}
\usepackage{multirow}

\DeclareFloatingEnvironment[name=Formulation]{formulation}

\def\BibTeX{{\rm B\kern-.05em{\sc i\kern-.025em b}\kern-.08em
    T\kern-.1667em\lower.7ex\hbox{E}\kern-.125emX}}

\makeatletter
\def\ps@IEEEtitlepagestyle{%
  \def\@oddhead{\mycopyrightnotice}%
  \def\@evenhead{}%
}
\def\mycopyrightnotice{%
  {\footnotesize\parbox{\textwidth}{This manuscript has been accepted for presentation at IEEE CCNC 2026. You can use this material personally. Reprinting or republishing this material for the purpose of advertising or promotion, creating new collective works, reselling or redistributing to servers or lists, or using any copyrighted component in other works must adhere to IEEE policy. The DOI will be supplied as soon as it becomes available.}}%
  \gdef\mycopyrightnotice{}%
}
\makeatother

\begin{document}

\title{Rethinking HTTP API Rate Limiting: A Client-Side Approach\\
\thanks{This work was supported by NSF CNS Award 2213672.}
}
\author{
    \IEEEauthorblockN{Behrooz Farkiani, Fan Liu, Patrick Crowley}\\
    \IEEEauthorblockA{
        Washington University in St. Louis, \\ 1 Brookings Dr., St. Louis, MO, 63130, USA\\
        Emails: \{b.farkiani, fan.liu, pcrowley\}@wustl.edu
    }
}

\maketitle

\begin{abstract}
HTTP underpins modern Internet services, and providers enforce quotas to regulate HTTP API traffic for scalability and reliability. When requests exceed quotas, clients are throttled and must retry. Server-side enforcement protects the service. However, when independent clients' usage counts toward a shared quota, server-only controls are inefficient; clients lack visibility into others' load, causing their retry attempts to potentially fail. Indeed, retry timing is important since each attempt incurs costs and yields no benefit unless admitted. While centralized coordination could address this, practical limitations have led to widespread adoption of simple client-side strategies like exponential backoff. As we show, these simple strategies cause excessive retries and significant costs. We design adaptive client-side mechanisms requiring no central control, relying only on minimal feedback. We present two algorithms: ATB, an offline method deployable via service workers, and AATB, which enhances retry behavior using aggregated telemetry data. Both algorithms infer system congestion to schedule retries. Through emulations with real-world traces and synthetic datasets with up to 100 clients, we demonstrate that our algorithms reduce HTTP 429 errors by up to 97.3\% compared to exponential backoff, while the modest increase in completion time is outweighed by the reduction in errors.
\end{abstract}

\begin{IEEEkeywords}
HTTP API, Rate Limiting, Distributed Algorithm, Congestion Control
\end{IEEEkeywords}
\section{Introduction}

Across the Internet, mechanisms at various network layers prevent senders from overwhelming receivers \cite{flach_internet-wide_2016, htb, saeed_carousel_2017, wu_traffic_2002, fu_survey_2023, jiang_when_2021}. At the application layer, HTTP carries over 50\% of Internet traffic and is central to service delivery \cite{tsareva_decade_2023, farkiani_hermes_2024}. HTTP APIs are ubiquitous in modern cloud services. To protect backend services, incoming HTTP traffic must be regulated \cite{el_malki_impact_2022}; as with other layers, controls can be deployed on both clients and servers. On the server side, providers must define quotas and monitor requests over time to prevent overload, which requires three elements: (1) mapping each request to its service usage, (2) selecting enforcement granularity and time horizons, and (3) enforcing quotas and handling excess usage \cite{api_pattern}.

The mapping from a client request to consumed service depends on the nature of service, ranging from one-to-one to cost-weighted calculations that reflect processing complexity (e.g., \cite{googlemap_rates, github_graphql, openai_rates}). Providers then choose enforcement granularity (IP address, API key, user ID) and interval (per second, minute, hour) to fit operational needs and set limits accordingly. Enforcement can use sliding window, leaky bucket, or token bucket \cite{alibaba_ratelimit} algorithms among others: sliding window counts requests within a time frame; the leaky bucket algorithm employs a fixed bucket size and processes requests at a constant rate, discarding any requests that exceed capacity; token bucket issues tokens at a fixed rate, accumulates unused tokens up to a burst capacity, and admits requests only when enough tokens are available. Notable examples include NGINX's leaky bucket \cite{nginx_ratelimit} and Envoy's token bucket \cite{envoy_token}. When a client exceeds its quota, services typically reject requests with HTTP 429 or HTTP 503 and may include helper headers such as \texttt{Retry-After} and \texttt{RateLimit-Limit} \cite{http_header}. Providers can also block IPs or apply other defenses \cite{api_pattern}. In most implementations, rejected requests count toward quotas to deter misuse (e.g., OpenAI\cite{openai_exponential}).

Server-side rate limiting protects services from excess traffic but can be inefficient when independent clients share a common quota. This situation arises, for example, at public, Internet-facing endpoints such as search or login pages \cite{cloudflare_rate}, or when multiple service instances hit the same rate-limited component \cite{skalski_leveraging_nodate}. In these cases, relying only on server-side mechanisms leaves each client guessing about others' behavior when deciding to retry, which prevents efficient use of available capacity.

Centralized solutions help but introduce new problems. Central proxies that queue requests from all clients and forward them to the rate-limited service (e.g., \cite{xu_rate_nodate, elnikety_method_2004}) simply move the bottleneck to the proxy as it needs to be protected from overloading. Central coordinators that schedule which client may send and when (e.g., \cite{starnberger_adaptive_2011, urgaonkar_cataclysm_2005}) incur high computational and communication overhead, depend on cooperation from the rate-limited service, and often do not scale to fine-grained scheduling. These limitations make client-side strategies appealing. The most common is exponential backoff, widely recommended in practice (e.g., Amazon \cite{amazon} and OpenAI \cite{openai_exponential}). Despite its simplicity, this approach leads to excessive retry attempts as our evaluation shows.

In this paper, we design client-side algorithms that use minimal information and do not require centralized coordinators, extensive server-side feedback, or central proxies. Our methods regulate retries by inferring congestion and adapting retry timing, rather than relying only on time as in exponential backoff and its variants. To the best of our knowledge, this is the first work to introduce non-time-based client-side algorithms that improve end-user experience when accessing rate-limited HTTP APIs. We ask a central question: can efficient non-time-based algorithms that use only minimal available information be used on the client side to improve service delivery? Our contributions are as follows:

\begin{enumerate}
    \item Model a centralized rate-limiting problem using mixed-integer linear programming (MILP).
    \item Propose two adaptive client-side algorithms, ATB and AATB, to enhance service delivery.
    \item Evaluate these algorithms using real-world and synthetic traces through emulation, showing that they reduce the number of HTTP~429 errors by more than 90\%.
\end{enumerate}

We organize the paper as follows: Section \ref{problem} states the problem, Section \ref{solution} presents the solutions, Section \ref{emulation} reports the evaluation, and Section \ref{conclusion} concludes.

\section{Problem Description}\label{problem}
This section formulates the \textit{rateLimiting} problem by considering an oracle that knows all requests in advance. 
\subsection{Assumptions}
We assume independent HTTP clients (for example, web browsers or custom clients) access a single HTTP endpoint, and their usage aggregates toward a common quota. Clients can buffer outgoing requests and send them in FIFO order to the endpoint that performs a function, which may be operated by a third party and is protected by a rate limiter. Examples include public endpoints such as login or search pages \cite{cloudflare_rate}, service-mesh endpoints used for inter-service communication \cite{isito_rate_limit}, and third-party APIs that require an API key, assuming authenticated clients already possess the necessary keys. Because clients share one quota, we assume equal priority across requests and clients, although the formulation can be extended to multiple priority classes that share a quota. For the sake of formulation, we assume the rate limiter uses a token bucket, but the specific algorithm is immaterial to our approach because our methods operate on the client side.

We assume client load can exceed the rate limiter's capacity and focus on client-side algorithms that decide when to send and, on failure, when to retry. Rejected attempts incur a cost that makes brute-force token acquisition impractical; the cost may be implicit (energy or network overhead) or explicit, as in some services (e.g., OpenAI \cite{openai_exponential}). Each request consumes one token, and once the quota is exhausted the limiter returns HTTP~429 until new quota becomes available. We further assume no IP blocking or additional restrictions are in place, and no feedback or helper headers beyond HTTP~429 responses are provided. This worst-case, minimal-information setting makes our solutions broadly applicable.

\subsection{Problem Formulation}
\begin{table}[ht]
\centering
\caption{Notations}
\label{tab:notation}
\begin{tabular}{ll}
\toprule
\multicolumn{2}{c}{\textbf{Parameters}} \\
\midrule
\textbf{Symbol} & \textbf{Description} \\
\(I\)       & Set of users  \\
\(J(i)\)    & Set of requests for user $i \in I$  \\
\(T\)       & Set of time slots, \(T=\{0,1,\ldots,T_{\max}\}\)  \\
\(B\)       & Token-bucket capacity and initial tokens  \\
\(r\)       & Token generation rate per time slot \\
\(A_{i,j}\) & Arrival time of request $(i, j), i \ \in I, j \in J(i)$ \\
\midrule
\multicolumn{2}{c}{\textbf{Decision Variables}} \\
\midrule
\textbf{Symbol} & \textbf{Description} \\
\(z_{i,j,t}\) & 1 if request \((i,j)\) is served at time \(t\), \(\in \{0,1\}\) \\
\(x_{i,j}\)   & Service time for request \((i,j)\), \(\in \mathbb{Z}^+\) \\
\(y_t\)       & Tokens available at the end of time slot \(t\), \(\in \mathbb{R}_{\ge 0}\) \\
\bottomrule
\end{tabular}
\end{table}
\begin{table}[htbp]
\centering
\begin{tabular}{@{}l p{0.9\linewidth}@{}}
\multicolumn{2}{c}{$\displaystyle \min \sum_{i \in I} \sum_{j \in J(i)} \bigl(x_{i,j} - A_{i,j}\bigr)$} \\
(1) & $\displaystyle
\sum_{t = \lceil A_{i,j} \rceil}^{T_{\max}} z_{i,j,t} = 1,
\quad \forall\, i \in I,\ \forall\, j \in J(i)
$ \\
(2) & $\displaystyle
\sum_{\tau = \lceil A_{i,j+1} \rceil}^{t} z_{i,j+1,\tau}
\le
\sum_{\tau = \lceil A_{i,j} \rceil}^{t} z_{i,j,\tau}
$ \\
& \quad $\forall\, i \in I,\ j \in J(i),\ t \in T,\ t \ge \lceil A_{i,j+1} \rceil$ \\
(3) & $\displaystyle
Z_t = \sum_{i \in I} \sum_{j \in J(i)} z_{i,j,t} \quad \forall\, t  \in T
$ \\
(4) & $\displaystyle
y_t = \min\bigl(y_{t-1} - Z_t + r,\ B\bigr)
$ \\
& \quad $\forall\, t = 1, \dots, T_{\max} \quad \text{with } y_0 = B$ \\
(5) & $\displaystyle
x_{i,j} = \sum_{t = \lceil A_{i,j} \rceil}^{T_{\max}} t\, z_{i,j,t},
\quad \forall\, i \in I,\ j \in J(i)
$ \\
\end{tabular}
\caption{The \textit{rateLimiting} problem.}
\label{form:rateLimiting}
\end{table}
Under the above assumptions and given an oracle that knows all requests in advance, the \textit{rateLimiting} problem is formulated as shown in Table \ref{form:rateLimiting}. The objective is to minimize the time elapsed between a request's arrival and when it is served (response time). Notations are shown in Table \ref{tab:notation}.

Constraints (1) and (2) ensure each request is accepted exactly once after its arrival and they are served in FIFO order. Constraint (3) defines the number of tokens consumed at each time, while Constraint (4) models the token bucket dynamics, including refill and capacity limits, and can be easily linearized. Constraint (5) defines the actual service time of each request.

In practice, requests are not known a priori. Therefore, minimizing the objective requires each node to brute-force sending requests to acquire tokens as soon as they become available. This leads to the highest cost and should be avoided. Next, we investigate practical client-side approaches.

\section{Solutions}\label{solution}

A common client-side approach is exponential backoff (e.g., OpenAI and Amazon \cite{amazon, openai_exponential}): a client sends a request as soon as it arrives; on failure, it waits for a random duration before retrying and doubles the upper bound of the wait after each failure. The upper bound is usually capped, but the client continues retrying until served. We call this \textit{Unlimited Backoff (UB)}. Variants of UB add controls to curb retries. For example, in \textit{Window-Based Backoff (WB)}, a sliding window restricts a client to at most \textit{W} requests (original and retries) per 60 seconds. On failure, the client applies exponential backoff, but after the timer expires, it must also verify the window limit; if the limit is exceeded, it waits until the window permits another request.

We can view the problem from a congestion control perspective. In TCP, a shared bottleneck link limits the rate of packet delivery and drops packets when the incoming rate exceeds its capacity. Similarly, the rate limiter functions as a shared bottleneck, dropping requests that exceed its quota. However, unlike TCP where a continuous stream of packets enables the detection of network changes, here we only observe dropped requests. We adopt a similar congestion-control approach by having each client implement an adaptive token bucket that permits sending only when a token is available. If a request is successfully served, the client increases its token generation rate, inferring that congestion is low. Conversely, if a request fails, the client records the congestion rate and decreases its token generation rate. We refer to this solution as the \textit{Adaptive Token Bucket (ATB)} algorithm.

This solution is straightforward and can be easily implemented in web browsers using a service worker, which acts as a proxy and can modify requests \cite{service_worker}. The pseudocode for this solution is shown in Table~\ref{alg:tb}. In the algorithm, $\sigma$ and $\delta$ are fixed parameters representing the minimum token-generation rate and the minimum increment to the token-generation rate, respectively. The tunable parameters $\alpha$ and $\beta$ determine how much the algorithm can increase the rate. The service provider initializes the token bucket capacity, the initial token count, the initial rate, and the congestion rate of each client. Each time a client sends a request, it needs to acquire a token. If the request is successful, it calls \texttt{INCREASE\_RATE}; if it fails, \texttt{DECREASE\_RATE} resets the token count and updates the token generation rate.
\begin{table}[htbp]
\centering
\caption{Adaptive Token Bucket Algorithm}
\label{alg:tb}
\vspace{-0.3cm} 
\begin{tabular}{@{}l@{}} 
\begin{minipage}{\columnwidth}
\begin{algorithmic}[1]
\Procedure{Acquire}{}
\State $tokens \gets \min(bucket\_size,\ tokens + (current\_time - last\_used)\times rate)$
\If{$tokens \geq 1$}
\State $tokens \gets tokens - 1$
\State $last\_used \gets now$
\State \textbf{return} success
\Else
\State wait until $tokens \geq 1$
\State $tokens \gets tokens - 1$
\State $last\_used \gets now$
\State \textbf{return} success
\EndIf
\EndProcedure

\Procedure{Increase\_rate}{}
\If{$rate < last\_congestion\_rate$}
\State $rate \gets \max(rate + \delta,\, rate \times \alpha)$
\Else
\State $rate \gets \max(rate + \delta,\, rate \times \beta)$
\EndIf
\EndProcedure

\Procedure{Decrease\_rate}{}
\State $last\_congestion\_rate \gets rate$
\State $tokens \gets 0$
\State $rate \gets \max(\sigma + \operatorname{rand}(-0.5,0.5),\, rate/2)$
\EndProcedure
\end{algorithmic}
\end{minipage}
\end{tabular}
\end{table}

Our second approach uses aggregated telemetry data to better infer congestion levels. We assume each client periodically sends data about the number of requests sent during the past $\omega$ seconds and whether they received HTTP 429 errors to a telemetry server over UDP. The telemetry server then aggregates data and informs clients about the current number of active clients, the total number of requests sent by all clients, the number of clients that received a 429, and the current rate limiter rates if it has been updated. Please note that this information does not require cooperation from the rate-limited service, and thus, we can apply this solution even when service is provided by a third-party. In addition, feedback from telemetry server may not be available at all times or may be provided selectively. Also, this feedback neither counts toward quotas nor serves as coordination messages. We refer to this solution as the \textit{Assisted Adaptive Token Bucket (AATB)} algorithm, as shown in Table~\ref{alg:assistedATB}, where $\omega$ denotes the report interval.

Unlike ATB, the client does not increase its rate with each successful request; instead, it routinely updates its rate based on the information received from the telemetry server. Before running the \texttt{ROUTINE\_UPDATE} procedure, the client first checks whether at least $\omega$ seconds have passed since the last reported congestion. Then, it sends telemetry data and receives updated information from the telemetry server. If any client reported HTTP 429 responses, indicating congestion during the past window, the client calculates \texttt{next\_acquire} to delay its next token acquisition. Otherwise, it compares its current load to the average load of all clients and adjusts its token generation rate accordingly.

If a client experiences a 429 error, it immediately notifies the network by calling the \texttt{CONGESTION\_NOTIFICATION} function. The client then uses the received data to reduce its rate by comparing its load to the average load of other nodes, and it updates \texttt{next\_acquire} based on the estimated duration needed for the current congestion to clear. The \texttt{ACQUIRE} function in AATB is similar to that in the Adaptive Token Bucket algorithm, with one key difference: when a token is available, the client also verifies that the current time is past \texttt{next\_acquire}; if not, or if no token is available, the client waits until both conditions are satisfied. 

\begin{table}[htbp]
\centering
\caption{Assisted Adaptive Token Bucket Algorithm}
\label{alg:assistedATB}
\vspace{-0.3cm} 
\begin{tabular}{@{}l@{}} 
\begin{minipage}{\columnwidth}
\begin{algorithmic}[1]

\Procedure{routine\_update}{}
\State Send telemetry data.
\If{$reported\_429 > 0$}
\State $backoff \gets \omega + \operatorname{rand}(-2, 2)$.
\State $next\_acquire \gets now + backoff$.
\ElsIf{$(now - last\_rate\_change) \geq \omega $}
\State Calculate average network and client load.
\If{$client\_load < 0.75 \times avg\_load$} 
\State $rate \gets \max\bigl(rate \times \alpha,\, rate + \delta\bigr)$.
\Else
\State $rate \gets \max\bigl(rate \times \beta,\, rate + \delta\bigr)$.
\EndIf
\State $last\_rate\_change \gets now$.
\EndIf
\EndProcedure

\Procedure{congestion\_notification}{}
\State $last\_reported\_congestion \gets now$.
\State Send telemetry data.
\State \textbf{Calculate} average network load and client load.
\If{$client\_load < 0.5 \times avg\_load$}
\State $new\_rate \gets \max\bigl(\sigma,\, rate/2\bigr)$.
\Else
\State $new\_rate \gets \max\bigl(\sigma,\, rate/3\bigr)$.
\EndIf
\State $rate \gets new\_rate$.
\State $tokens \gets 1.1$ \Comment{Allow immediate send.}
\State $last\_rate\_change \gets now$.
\State $wait\_time \gets \frac{reported\_429}{token\_rate} +\operatorname{rand}(0,1)$.
\State $next\_acquire \gets now + wait\_time$.
\EndProcedure
\end{algorithmic}
\end{minipage}
\end{tabular}
\end{table}

\section{Implementation and Evaluation}\label{emulation}

\begin{figure*}[htbp]
\centering
\begin{subfigure}[b]{0.3\textwidth}
\includegraphics[width=\textwidth]{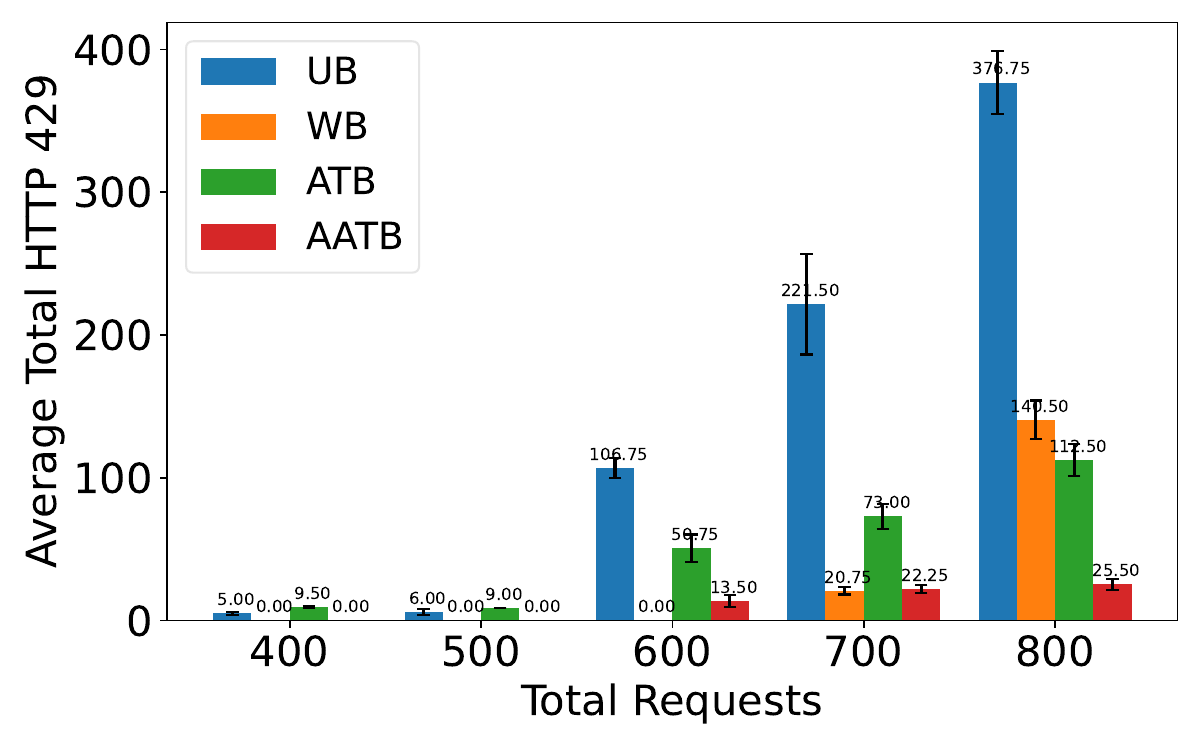}
\caption{Real trace - HTTP errors}
\label{fig:test_a_1}
\end{subfigure}
\hfill
\begin{subfigure}[b]{0.3\textwidth}
\includegraphics[width=\textwidth]{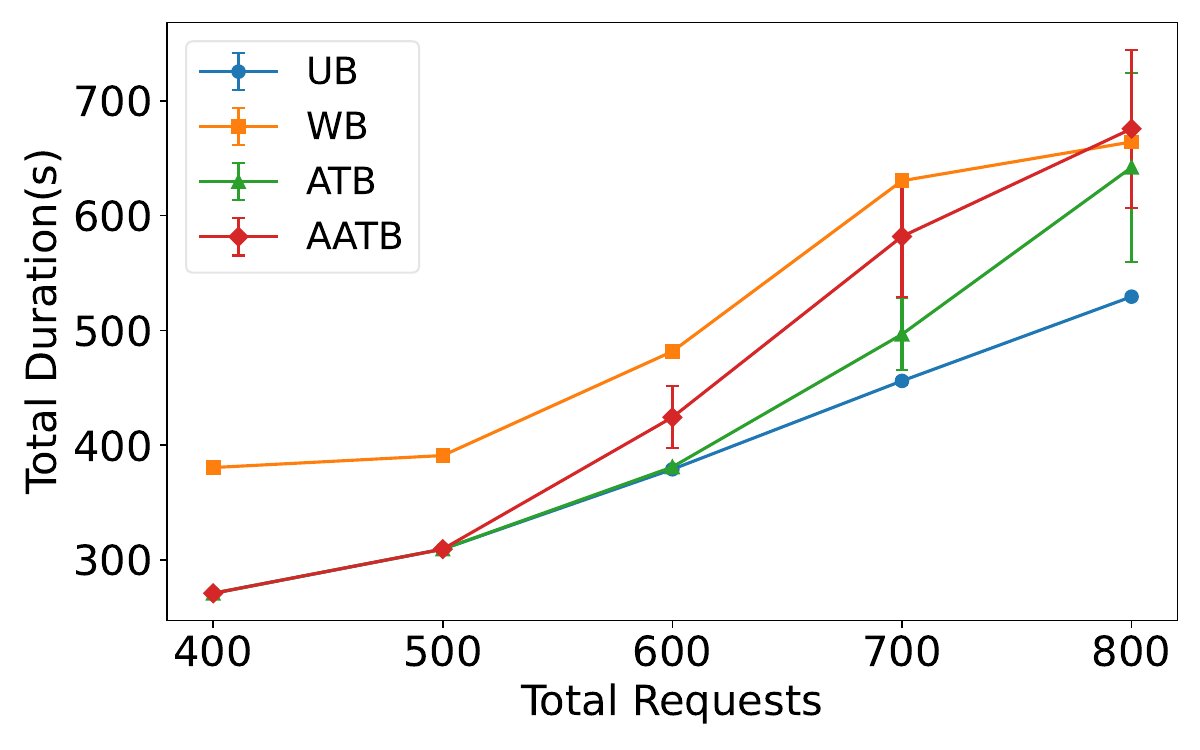}
\caption{Real trace - Duration}
\label{fig:test_b_1}
\end{subfigure}
\hfill
\begin{subfigure}[b]{0.3\textwidth}
\includegraphics[width=\textwidth]{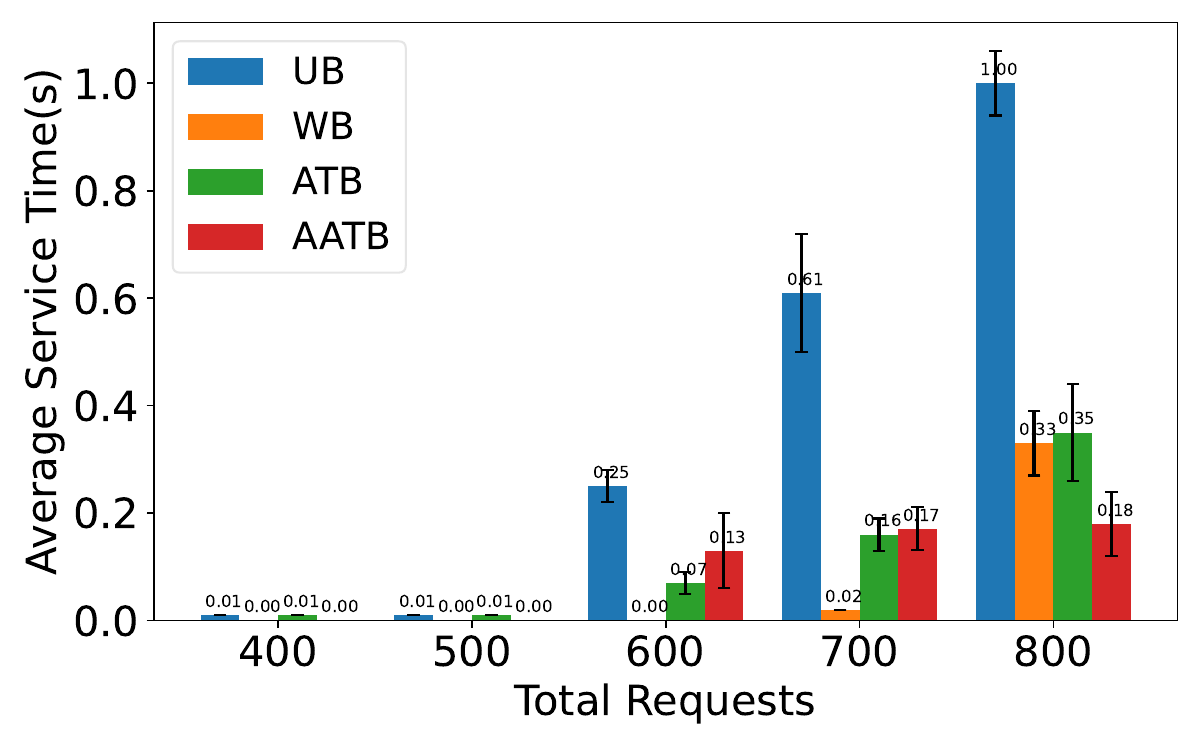}
\caption{Real trace datasets - Service time}
\label{fig:test_c_1}
\end{subfigure}


\begin{subfigure}[b]{0.3\textwidth}
\includegraphics[width=\textwidth]{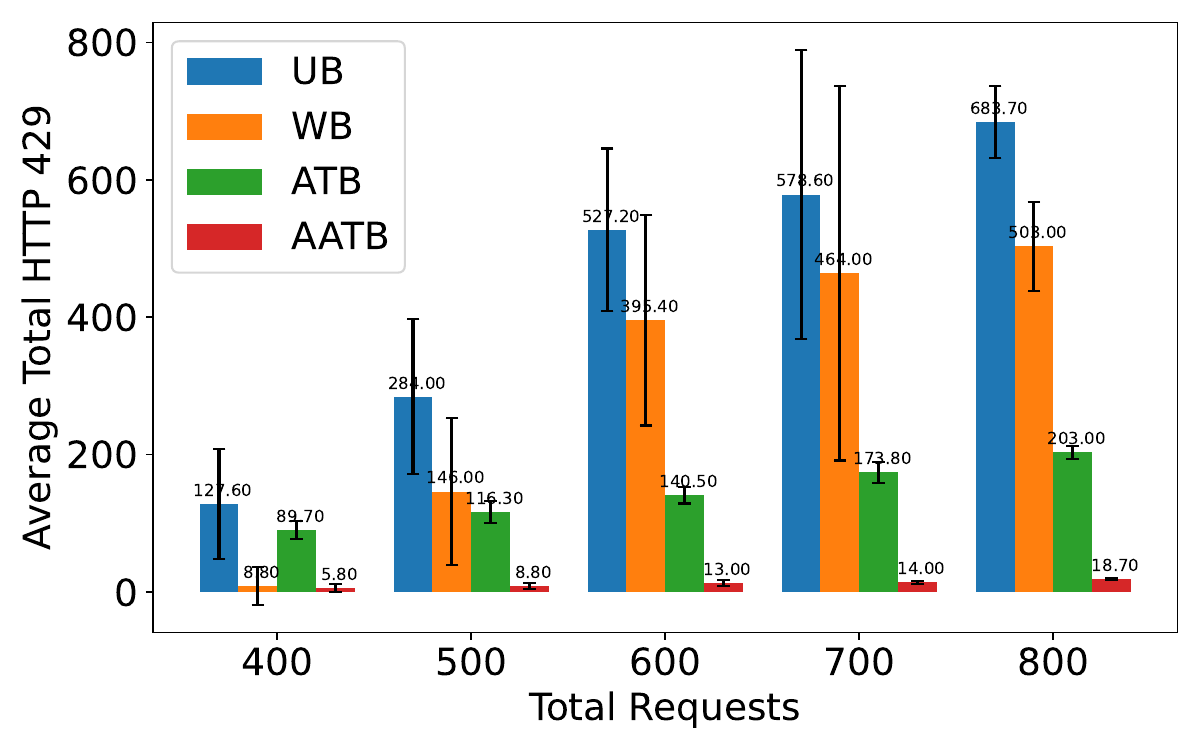}
\caption{Five-client - HTTP errors}
\label{fig:fivCa}
\end{subfigure}
\hfill
\begin{subfigure}[b]{0.3\textwidth}
\includegraphics[width=\textwidth]{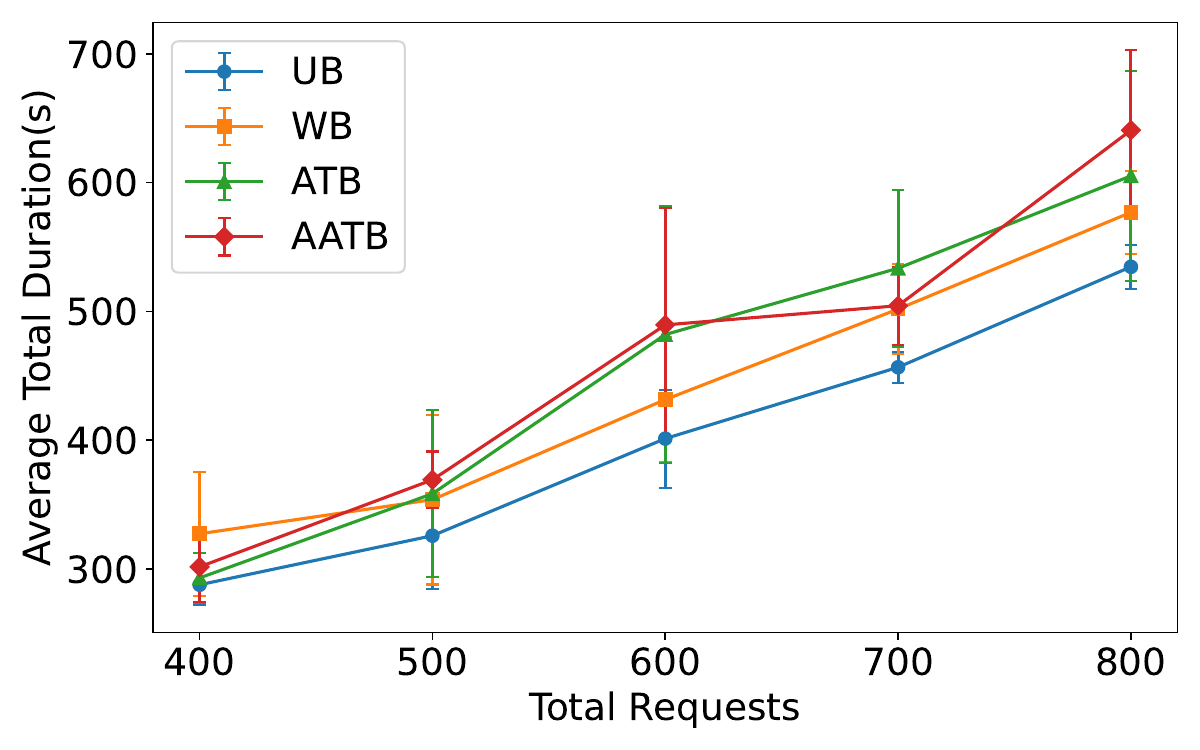}
\caption{Five-client - Duration}
\label{fig:fivCb}
\end{subfigure}
\hfill
\begin{subfigure}[b]{0.3\textwidth}
\includegraphics[width=\textwidth]{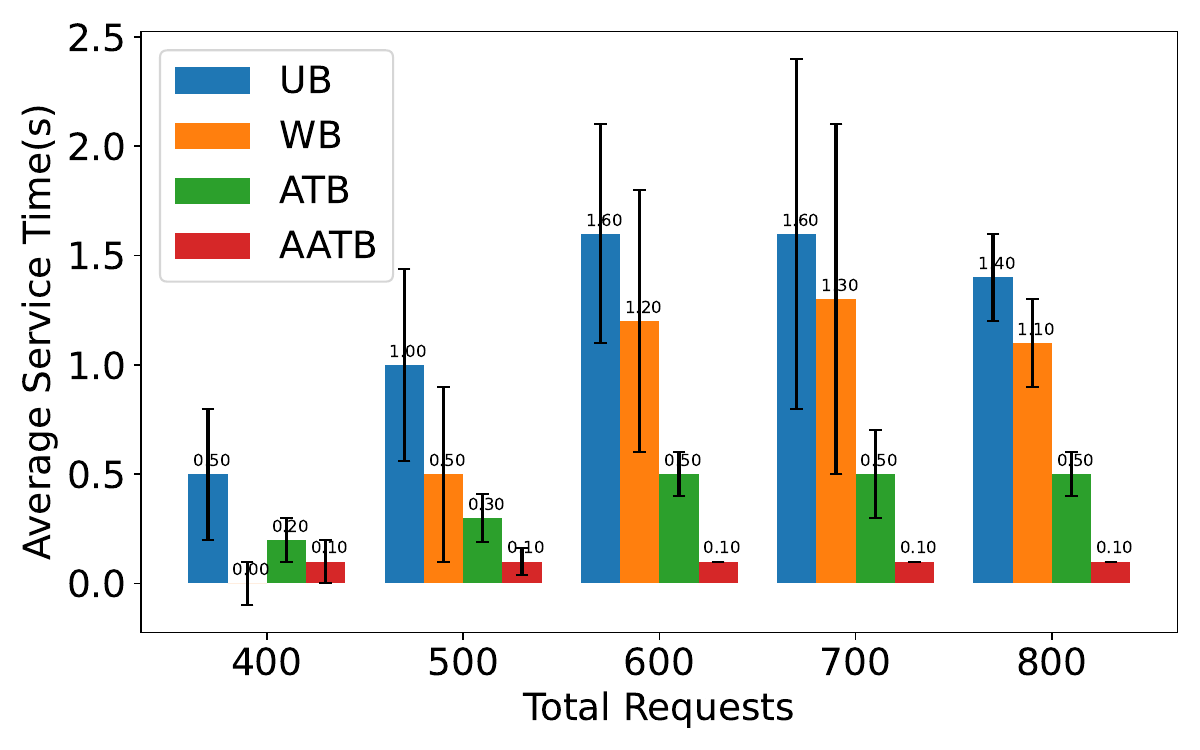}
\caption{Five-client - Service time}
\label{fig:fivCc}
\end{subfigure}


\begin{subfigure}[b]{0.3\textwidth}
\includegraphics[width=\textwidth]{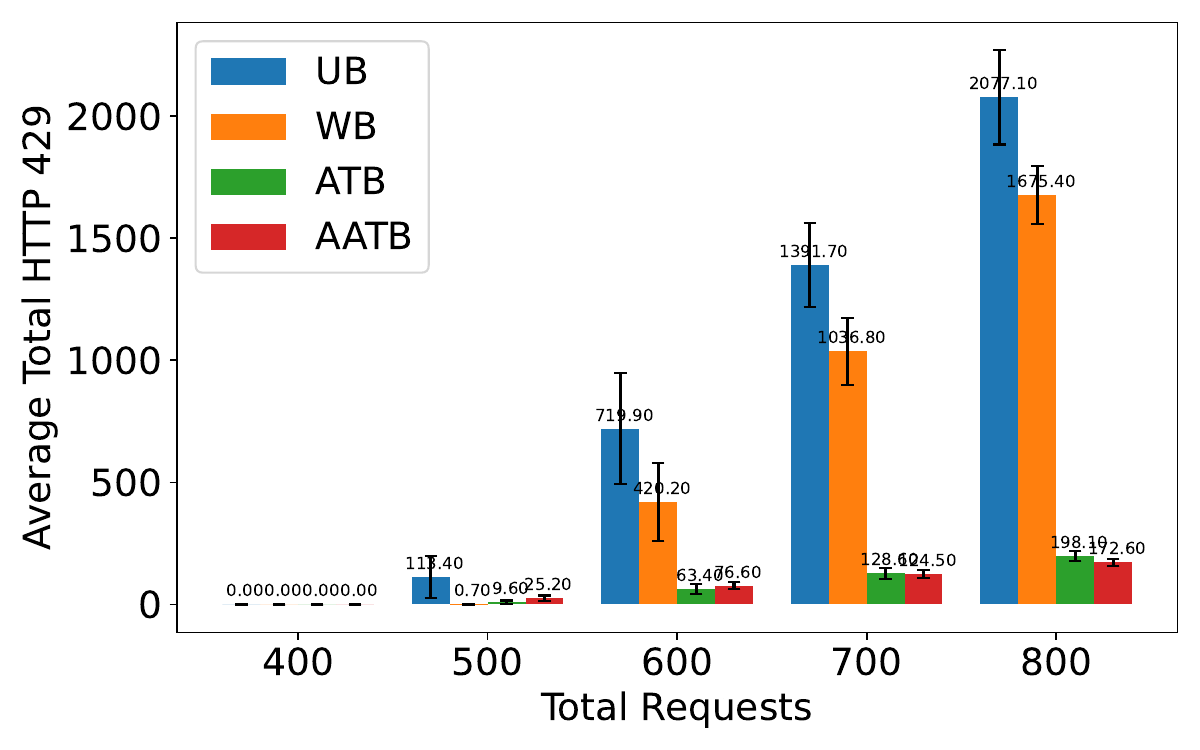}
\caption{One-hundred-client - HTTP errors}
\label{fig:hunCa}
\end{subfigure}
\hfill
\begin{subfigure}[b]{0.3\textwidth}
\includegraphics[width=\textwidth]{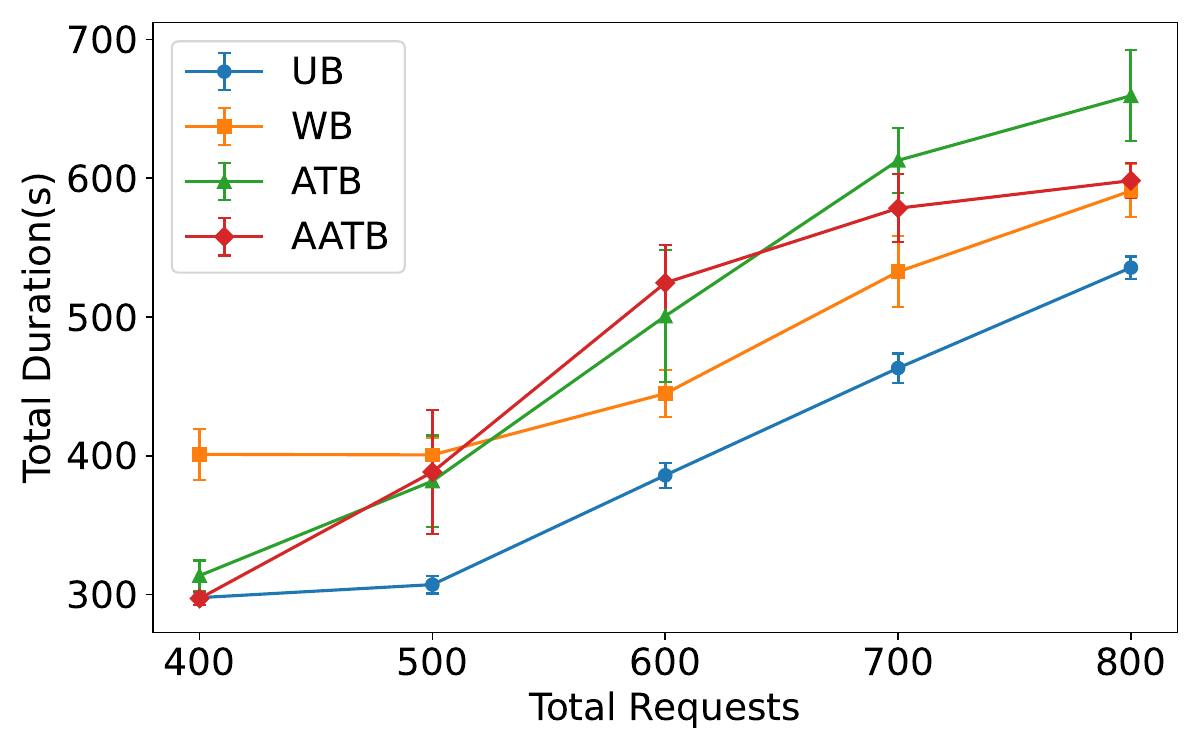}
\caption{One-hundred-client - Duration}
\label{fig:hunCb}
\end{subfigure}
\hfill
\begin{subfigure}[b]{0.3\textwidth}
\includegraphics[width=\textwidth]{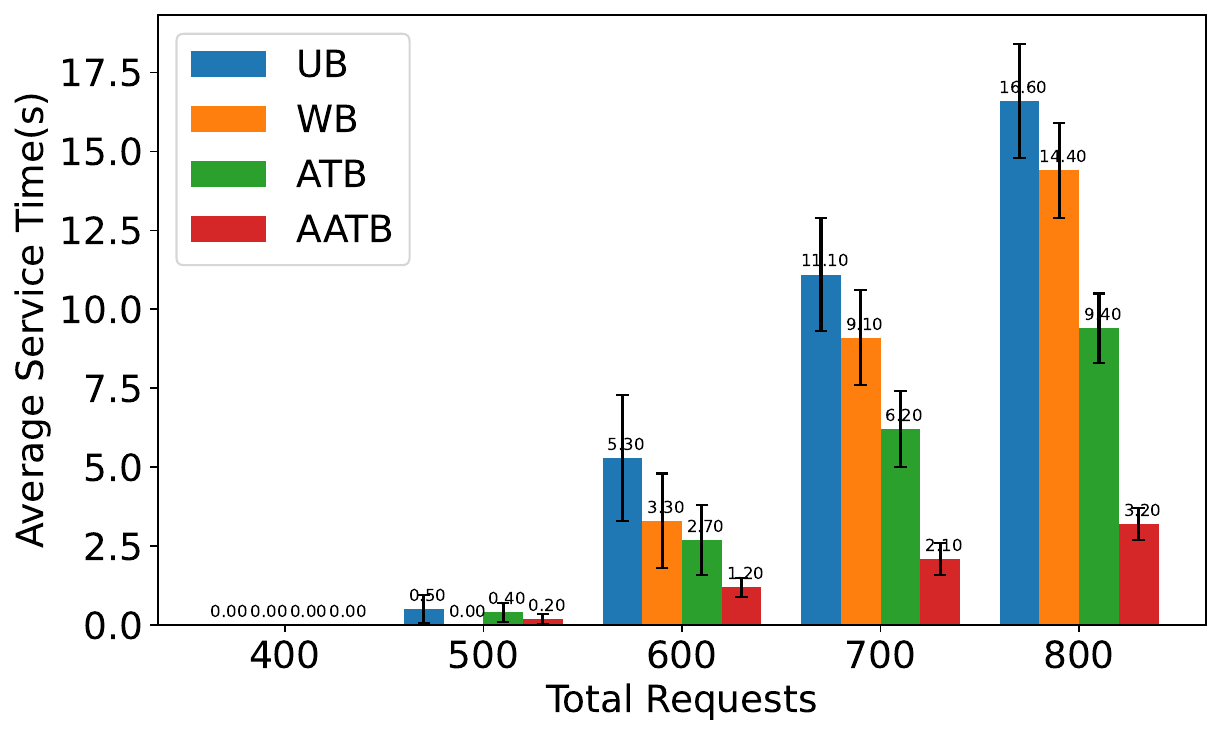}
\caption{One-hundred-client - Service time}
\label{fig:hunCc}
\end{subfigure}

\caption{Evaluation results: (a-c) Real-world trace, (d-f) Five-client scenario, (g-i) One-hundred-client scenario. Standard deviation is shown as error bars.}
\label{fig:combined_results}
\end{figure*}

We emulated varying client counts and workloads using two virtual machines on a single physical host. The telemetry server, application service, and clients were implemented in Python 3.13 with \texttt{asyncio}. Clients started asynchronously with random delays up to 10s and ran in separate processes with no inter-client communication. In addition to the Python clients, we built and tested a service worker for ATB and a reference implementation of the \textit{rateLimiting} problem using the CPLEX Python API\footnote{All implementations and datasets are available at \url{https://github.com/Bfarkiani/ratelimiter}}. The server used Hypercorn \cite{jones_pgjoneshypercorn_2025} with HTTP/2 over cleartext (h2c), a backlog of 2048, and a keep-alive timeout of 350s. Each HTTP API request carried two fixed numbers in JSON, and the service, fronted and rate-limited by Envoy v1.33.0 \cite{envoy}, returned their product. This minimal service logic reduces server-side variability, so once requests pass the Envoy rate limiter they incur no additional server-side delay. 

To evaluate the algorithms, we used real-world \cite{lagopoulos_web_2019} and synthetic traces. The real-world trace is a search-endpoint access log that averages 131K requests per day and 27K unique IP addresses. We mapped each IP address to a user and split the data into training and test sets, then constructed datasets of sizes 400 to 800 requests in steps of 100; the training sets were used to tune parameters of WB, ATB, and AATB via a simple parameter search. Details of the test sets appear in Table~\ref{tab:test}. Because the number of users varied in the real trace, we also generated synthetic traces to remove this variability. The synthetic evaluation considered two scenarios: (1) a five-client case that represented a service mesh with five high-traffic services, and (2) a one-hundred-client case that resembled an endpoint with many clients, similar to the real trace. For synthetic datasets, we generated 400 to 800 requests within a 5-minute interval. Each client was assigned one request; additional requests were sampled from a Poisson distribution with parameter $\lambda=\text{range}/2$ (range in Table~\ref{tab:test_param}), and timestamps come from an exponential distribution with scale $\lambda$. For all evaluations, Envoy was configured in front of the server as the rate limiter with a token-bucket capacity of 100 and a token generation rate of 80 tokens per minute. This admitted 500 requests in a 5-minute experiment, which corresponded to 144K requests per day. In AATB, all clients send telemetry data and the telemetry server always responds. All synthetic results were averaged over at least 30 runs. We report the following metrics, averaged over all clients and runs, to measure algorithm performance.
\begin{itemize}
\item \textbf{Average total emulation duration:} The time from when the first request is generated until the last request is served among all clients, including waiting time in the client buffer.
\item \textbf{Average service time:} The duration from when a request leaves the client queue for the first time until it is served, accounting for retries.
\item \textbf{Average total number of 429 errors:} The total number of HTTP 429 errors received by all clients.
\end{itemize} 

\begin{table}[ht]
\centering
\caption{Algorithm configurations}
\label{tab:test_param}
\begin{tabular}{lp{0.65\linewidth}}
\toprule
\multicolumn{2}{c}{\textbf{Real datasets}} \\
\midrule
UB & Backoff upper bound $\sim$ Uniform[30, 34] \\
WB & W $\leftarrow$ 15 per minute \\
ATB & $\sigma \leftarrow 0.6$/minute, $\delta \leftarrow 0.6$/minute, initial tokens $\leftarrow$ 1, $\alpha \leftarrow 1.2$, $\beta \leftarrow 1.2$ \newline
initial congestion rate $\leftarrow$ 30 per minute,
bucket size $\leftarrow$ 15, initial rate $\leftarrow$ 15 per minute \\
AATB & Same as ATB with $\alpha \leftarrow 1.4$, $\beta \leftarrow 1.2$, $\omega \leftarrow 30$s \\
\midrule
\multicolumn{2}{c}{\textbf{Synthetic datasets, 5 and 100 for 5- and 100-client scenarios}} \\
\midrule
Requests range & \textbf{5}: range[1--200], \textbf{100}: range[1--10] \\
UB & Same as the real \\
WB & W $\leftarrow$ \textbf{5}:40, \textbf{100}:4 per minute \\
ATB & Same $\sigma, \delta$ ,  initial tokens, $\alpha, \beta$ as the real, \newline initial congestion rate $\leftarrow$ \textbf{5}:300, \textbf{100}:12 per minute,
bucket size $\leftarrow$ \textbf{5}:40, \textbf{100}:4, initial rate $\leftarrow$ \textbf{5}:40, \textbf{100}:4 per minute \\
AATB & Same as the real \\
\bottomrule
\end{tabular}
\end{table}

\begin{table}[ht]
\caption{Details of the test datasets from \cite{lagopoulos_web_2019}}
\label{tab:test}
\begin{tabular}{|l|lllll|}
\hline
\multicolumn{1}{|c|}{\multirow{2}{*}{\textbf{Attribute}}} & \multicolumn{5}{c|}{\textbf{Size}} \\ \cline{2-6} 
\multicolumn{1}{|c|}{} & \multicolumn{1}{c|}{400} & \multicolumn{1}{c|}{500} & \multicolumn{1}{c|}{600} & \multicolumn{1}{c|}{700} & \multicolumn{1}{c|}{800} \\ \hline
Number of users & \multicolumn{1}{l|}{18} & \multicolumn{1}{l|}{22} & \multicolumn{1}{l|}{23} & \multicolumn{1}{l|}{25} & 27 \\ \hline
Last time stamp(s) & \multicolumn{1}{l|}{263} & \multicolumn{1}{l|}{307} & \multicolumn{1}{l|}{339} & \multicolumn{1}{l|}{361} & 433 \\ \hline
\end{tabular}
\end{table}

Results for real-world test datasets are shown in Figures~\ref{fig:test_a_1} to \ref{fig:test_c_1}. As shown, UB has the highest number of errors. At 800 requests, WB reduces errors by 62.70\% with a 25.45\% increase in duration; ATB reduces errors by 70.13\% with a 21.26\% increase in duration, and AATB reduces errors by 93.23\% while increasing duration by 27.62\%. As the number of clients and load increases, AATB clients send more update messages and at 800 requests, they send an average of 276.25 update messages. Please note that these are telemetry messages and do not count toward quotas.

Figures~\ref{fig:fivCa} to \ref{fig:fivCc} show the five-client results. At 400 requests, all algorithms except UB and ATB exceed 300 seconds; at 500 requests, all exceed 300 seconds. UB finishes first but with more HTTP 429 errors and thus higher cost. Relative to UB, WB reduces errors by 48.6\% and 26.4\% with duration increases of 8.6\% and 7.9\% for 500 and 800 requests. ATB reduces errors by 59\% and 70.3\% with duration increases of 10\% and 13.2\%, and AATB reduces errors by 96.9\% and 97.3\% with duration increases of 13.3\% and 19.8\%, for 500 and 800 requests respectively. AATB also sends fewer than 88 update messages for all request sizes. As Figure~\ref{fig:fivCc} indicates, UB's random backoff gives the lowest total duration but the worst service time. ATB and AATB delay strategically, which raises total duration modestly but lowers service time and sharply cuts errors. Overall, the large drop in HTTP~429 errors outweighs the modest increase in duration for the proposed algorithms.

The one-hundred-client results are shown in Figures~\ref{fig:hunCa} to \ref{fig:hunCc} and mirror the five-client trends. At 400 requests, only UB and AATB finish within 300 seconds. At 800 requests, because there are 100 clients, the average number of AATB update messages is 1106; however, these are routine telemetry messages and do not count toward the quota. At 500 requests, WB reduces HTTP~429 errors by 99.4\% with a 30.4\% increase in total duration, but at 800 requests it cuts errors by only 19.3\% while raising duration by 10.3\%, which offers limited benefit. Algorithms are expected to perform well when the load exceeds the configured quotas. For 500 requests, ATB increases duration by 24.3\% and reduces errors by 91.5\%; at 800 requests, duration rises by 23.1\% and errors fall by 90.5\%, both relative to UB. AATB increases duration by 26.4\% for 500 requests and 11.7\% for 800 requests, while reducing errors by 77.8\% and 91.7\%, respectively, compared to UB. 

Without a central coordinator, one can drive errors to zero by delaying aggressively, although total duration becomes unacceptable. ATB and AATB target a better trade-off by inferring congestion level and strategically delaying sending requests, which also leads to better service time. Considering all results, ATB outperforms UB and WB, and AATB typically delivers the best results by sharply reducing errors while only modestly increasing duration. Therefore, ATB is an effective replacement for exponential backoff, whereas AATB performs better under heavy load conditions.

\section{Conclusion and Future Work}\label{conclusion}
This paper studied client-side algorithms for improving service delivery in rate-limited services when independent clients share a common quota. Using emulations with real traces and synthetic workloads, we found that commonly implemented exponential backoff algorithm produces many HTTP~429 errors causing significant costs, and its window-based variant's performance degrades under heavy load. We introduced two algorithms ATB and AATB that consider system congestion instead of solely relying on time. We showed that with datasets of 800 requests (up to 1.6 times the planned capacity), our algorithms reduced errors by 70.13\%-97.3\% with a 11.7\%-27.62\% increase in total duration. 

We designed these algorithms to use minimal information under worst-case assumptions for broad applicability. The large error reductions, often exceeding 90\%, justify the modest increases in duration. As future work, we will explore using lightweight helper headers to further reduce total duration and study how AATB update frequency affects performance.
\bibliographystyle{IEEEtran}
\bibliography{ref}

\end{document}